\documentclass[review,12pt]{elsarticle}

\usepackage{extsizes}
\usepackage[version=3]{mhchem}
\usepackage[left=2.5cm, right=2.5cm, top=1.785cm, bottom=2.0cm]{geometry}
\usepackage{times,mathptmx}
\usepackage{gensymb}
\usepackage{sectsty}
\usepackage{epsfig}
\usepackage{lastpage}
\usepackage{float}
\usepackage{fnpos}
\usepackage[english]{babel}
\usepackage{array}
\usepackage{droidsans}
\usepackage{charter}
\usepackage[T1]{fontenc}
\usepackage[usenames,dvipsnames]{xcolor}
\usepackage{setspace}
\usepackage[compact]{titlesec}
\usepackage{array}
\usepackage{color}
\usepackage{textcomp}
\usepackage{amsmath}
\usepackage{amssymb}
\usepackage{amsthm}

\usepackage{epstopdf}

\makeatletter
\def\ps@pprintTitle{%
     \let\@oddhead\@empty
     \let\@evenhead\@empty
     \def\@oddfoot{}%
		 \let\@evenfoot\oddfoot}
\makeatother

\begin{document}

\begin{frontmatter}

\title{Dynamical behavior of microgels of Interpenetrated Polymer Networks}

\author[DIP,ISC]{Valentina Nigro \footnote{Corresponding author: valentina.nigro@uniroma1.it}}
\author[ISC,DIP]{Roberta Angelini}
\author[IPCF]{Monica Bertoldo}
\author[RM3]{Fabio Bruni}
\author[RM3]{Maria Antonietta Ricci}
\author[ISC,DIP]{Barbara Ruzicka}

\address[DIP]{Dipartimento di Fisica, Sapienza Universit$\grave{a}$ di Roma, P.le Aldo Moro 5, 00185 Roma, Italy.}
\address[ISC]{Istituto dei Sistemi Complessi del Consiglio Nazionale delle Ricerche (ISC-CNR), Sede Sapienza, Pz.le Aldo Moro 5, 00185 Roma, Italy}

\address[IPCF]{Istituto per i Processi Chimico-Fisici del Consiglio Nazionale delle Ricerche (IPCF-CNR), Area della Ricerca, Via G.Moruzzi 1, I-56124 Pisa, Italy.}
\address[RM3]{Dipartimento di Scienze, Sezione di Nanoscienze, Universit$\grave{a}$  degli Studi Roma Tre, Via della Vasca Navale 84, I-00146 Roma, Italy}

\begin{abstract}
Microgel suspensions of Interpenetrated Polymer Network (IPN) of PNIPAM and PAAc in D$_2$O, have been investigated through dynamic light scattering as a function of temperature, pH and concentration across the Volume Phase Transition (VPT). The dynamics of the system is slowed down under H/D isotopic substitution due to the different balance between polymer/polymer and polymer/solvent interactions suggesting the crucial role played by H-bondings. The swelling behavior, reduced with respect to PNIPAM and water, has been described by the Flory-Rehner theory, tested for PNIPAM microgel and successfully expanded to higher order for IPN microgels. 
Moreover the concentration dependence of the relaxation time at neutral pH has highlighted two different routes to approach the glass transition: Arrhenius and super-Arrhenius (Vogel Fulcher Tammann) respectively below and above the VPT and a fragility plot has been derived. Fragility can be tuned by changing temperature: across the VPT particles undergo a transition from soft-strong to stiff-fragile.  
\\
\end{abstract}

\end{frontmatter}

%%%END OF FOOTNOTES%%%

%%%MAIN TEXT%%%%

\section{Introduction}
\label{Intro}

Research on colloidal systems has attracted great interest in the last years, due to the variety of their technological applications and to the richness of their phase behavior. Indeed they are very good model systems for understanding the general problem of dynamic arrest, since their larger tunability with respect to atomic and molecular systems~\cite{SciAdvPhys2005,TrappeCOCIS2004,PoonCOCIS1998,ZacJPCM2008} leads to complex phase diagrams, including different arrested states (such as gels ~\cite{LuNat2008, RoyallNatMat2008, RuzickaNatMat2011} and glasses ~\cite{PuseyNat1986, ImhofPRL1995}) and unusual glass-glass transitions ~\cite{PhamScience2002, EckertPRL2002, AngeliniNC2014}.

In particular among colloidal systems, soft colloids represent an interesting class of glass-formers, since, at variance with hardsphere-
like colloids, they are characterized by an interparticle potential with a finite repulsion at or beyond contact. As a result
of the particle softness a complex phase behavior has been theoretically predicted \cite{LikosJPCM2002,RamirezJPCM2009} and
not yet experimentally reproduced. 
Moreover, it has been recently shown \cite{MattssonNature2009} that soft colloids exhibit the same wide variation of the structural relaxation time observed for supercooled molecular liquids approaching the glassy state. 
This variation is characterized by the unifying concept of fragility, which gives a "universal" description of  dynamic arrest in glass-forming liquids \cite{AngellPNAS1995, AngellJAP2000, CasaliniJCP2012}. 
The relaxation time of "fragile" liquids  increases very rapidly with decreasing temperature, while "strong" liquids show a much slower temperature dependence. For colloidal suspensions fragility must be defined replacing the inverse of temperature with concentration \cite{MattssonNature2009, CasaliniJCP2012, McKennaPRE2016}. Hard-sphere colloidal suspensions are fragile and the absence of a wider range of fragility limits their versatility as a model system of the glass
transition. 
Interestingly, for deformable soft colloidal particles, fragility is affected by their elastic properties, giving rise to strong behavior \cite{AngellPNAS1995, SeekellSM2015} and allowing to obtain the equivalent effect of molecular systems \cite{AngellJAP2000}. Nevertheless, many efforts have been devoted to understanding the influence of the softness of the interparticle potential on the fragility of glass formers \cite{BordatPRL2004, SenguptaJCP2011, ShiJCP2011, CasaliniJCP2012}, with controversial results between computer simulations \cite{SenguptaJCP2011, ShiJCP2011} and experimental observations \cite{MattssonNature2009, McKennaJCP2014}. 

In this framework responsive microgels (aqueous suspensions of nanometre- or micrometre-sized hydrogel particles) allow to change their effective volume fraction and their elastic properties by tuning their response to an external stimulus. In particular they may exhibit high sensitivity to changes of pH, temperature, electric field, ionic strength, solvent or to external stresses or light pulses. 
These easily accessible external parameters allow to modulate the interparticle potential and their reversible Volume Phase Transition (VPT) (swelling/shrinking behavior), giving rise to novel phase-behaviors, drastically different from those of conventional hard-spheres-like colloidal systems \cite{ WangChemPhys2014, PritiJCP2014, HellwegCPS2000, PaloliSM2012, WuPRL2003}.

One of the most studied responsive microgel is based on a
 thermo-sensitive polymer, the  poly(N-isopropylacrylamide), also known as PNIPAM. Indeed PNIPAM-based microgels have been widely investigated in the last years~\cite{PeltonColloids1986, SaundersACIS1999, PeltonAdvColloid2000, DasAnnRevMR2006, KargCOCIS2009,
LuProgPolSci2011} and it is well known that their thermo-responsiveness is strongly
related to the coil-to-globule transition with temperature of NIPAM polymer.  At
room temperature indeed, the polymer is hydrophilic and strongly hydrated in solution, while it becomes hydrophobic above 305 K corresponding to its Lower Critical Solution Temperature. This gives rise to a Volume Phase Transition (VPT) from a swollen to a shrunken state of any microgel based on NIPAM polymer~\cite{WuMacromol2003}. Moreover it has been shown that this typical swelling/shrinking transition is the driving mechanism of the phase behavior of aqueous suspensions of PNIPAM microgels \cite{MattssonNature2009, PaloliSM2012, LyonRevPC2012, PritiJCP2014}. 
This typical swelling behavior can be strongly affected by
concentration~\cite{TanPolymers2010,WangChemPhys2014},
 solvents~\cite{ZhuMacroChemPhys1999} and
synthesis procedure (such as growing number of cross-linking
points~\cite{KratzBerBunsenges11998, KratzPolymer2001}, different
reaction pH conditions~\cite{BaoMacromol2006} or by introducing
additives into the PNIPAM network~\cite{HellwegLangmuir2004}).

In this context PNIPAM microgels containing another specie as
co\textendash{}monomer or interpenetrated polymer are
even more interesting, as a more complex scenario can show up. In particular, addition of
poly-acrilic acid (PAAc) to PNIPAM microgel provides pH-sensitivity to the thermo-responsive microgel. 
The VPT of these microgels strongly depends
on the effective charge density, controlled by the content of AAc
monomer~\cite{HuAdvMater2004, MaColloidInt2010}, on the pH of the
suspension~\cite{KratzColloids2000, KratzBerBunsenges21998,
XiaLangmuir2004, JonesMacromol2000, NigroJNCS2015, NigroJCP2015} and on salt
concentration~\cite{KratzColloids2000, XiongColloidSurf2011}.

This means that the synthesis procedure plays a crucial role, being the response of PNIPAM/PAAc microgels strictly related 
to the mutual interference between the two monomers \cite{XiongColloidSurf2011,
MengPhysChem2007, LyonJPCB2004, HolmqvistPRL2012, DebordJPCB2003}. 
While microgels made of random co-polymers (co-polymerised) have been widely investigated both experimentally and theoretically \cite{RomeoSM2013, RomeoJCP2012, KratzColloids2000, JonesMacromol2000, MengPhysChem2007}, a deep investigation of Interpenetrated Polymer Network (IPN) microgels  is still lacking. Interpenetration of hydrophilic PAAc 
into the PNIPAM microgels network (IPN PNIPAM-PAAc)~\cite{HuAdvMater2004, XiaLangmuir2004, XiaJCRel2005,
ZhouBio2008, XingCollPolym2010, LiuPolymers2012} provides independent sensitivity to temperature and pH, preserving the same VPT of pure PNIPAM microgel. 
Moreover the different solubility of PAAc at acidic and neutral pH introduces an additional control parameter that allows to tune the mutual interference between PNIPAM and PAAc networks. At acidic pH the carboxylic  (COOH-) groups of PAAc chains are protonated and H-bonds formation with neighbouring COOH- groups or with the amidic (CONH-) groups of PNIPAM in the same particle is favored, with respect to H-bonding with water molecules~\cite{SibandMacromol2011}. At neutral pH, the balance between PNIPAM/PAAc and water/PAAc H-bonds is reversed. Both compounds are therefore well solvated and water mediates their interaction, making the two networks completely independent one to each other. 

The swelling behavior of aqueous suspensions of PNIPAM-PAAc IPN microgels as a function of temperature, pH and concentration has been extensively investigated by our group and the obtained results have been published elsewhere \cite{NigroJNCS2015, NigroJCP2015, NigroCSA2017}. 

In order to understand the role played by hydrophobicity and hydrogen bonding in the polymer-solvent interactions and in the dynamics of the polymer solutions, investigations of the isotopic effect give relevant information. Indeed it has been recently shown that a slowing down of the swelling kinetics and a shift forward of the Volume Phase Transition Temperature (VPTT) of PNIPAM microgels occur in D$_2$O suspensions with respect to H$_2$O ones \cite{ShirotaMacromolSymp2004}, mainly due to the higher viscosity of the deuterated solvent. Similar results have been found also in other colloidal suspensions \cite{TudiscaRSC2012, MarquesSM2015, MarquesJPC2017}. 

In this work we present a DLS systematic investigation of the dynamics across the VPT of IPN microgels suspensions in
D$_2$O compared to H$_2$O as a function of temperature, pH and
concentration. Experimental data have been described through theoretical models
from the Flory-Rehner theory and discussed within the universal framework
of fragility.

\section{Experimental Methods}
\label{Experimental Methods}

\subsection{Sample preparation}

Once synthesized by a sequential free radical polymerization method, lyophilized IPN microgels were dispersed in H$_2$O or D$_2$O by magnetic stirring for 1 day.
The samples were then lyophilized and redispersed in H$_2$O or D$_2$O 
using the same molar ratio; the weight concentration, $C_w$,
reported in the text always refers to the case of H$_2$O.
Samples at different concentrations were obtained by dilution at two pH: pH 5, where H-bondings between PNIPAM and PAAc are favored, and pH 7, corresponding to the dissociation of the ionic groups and the reduction of the PNIPAM-PAAc H-bonds interactions. Samples at pH 7 were obtained by addition of NaOH or NaOD to samples at pH 5.
A detailed description of IPN microgel particles synthesis is reported in Ref.\cite{NigroJNCS2015, NigroJCP2015}.

 \subsection{DLS set-up and data analysis}
 \label{DLS set-up and data analysis}

\begin{figure}[t]
\centering
\includegraphics[height=8cm]{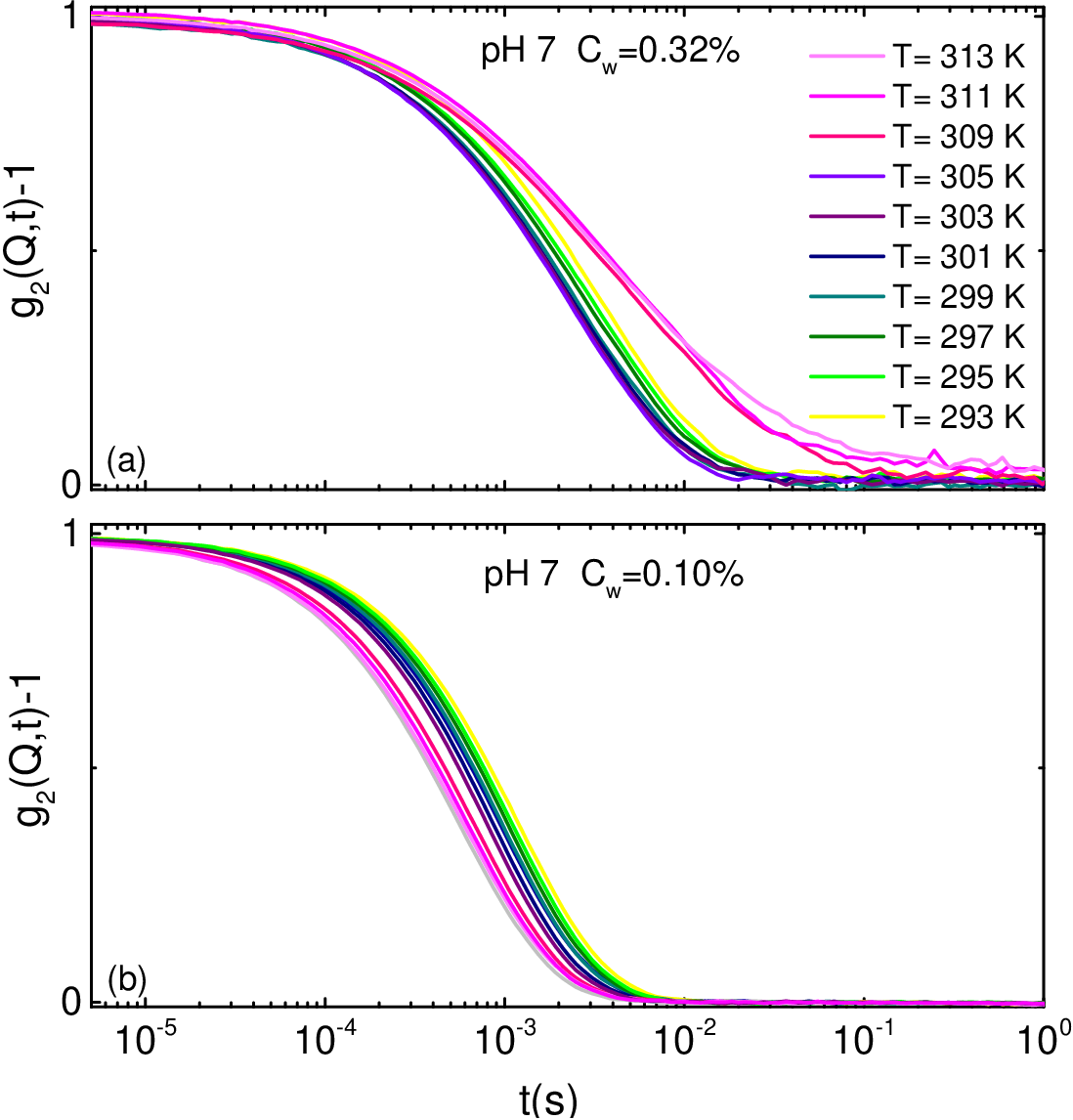}
\caption{Normalized intensity autocorrelation function of a D$_2$O suspension of IPN microgels at (a) $C_w$=0.32 \% and (b) $C_w$=0.10 \%, at pH 7 and Q=1.8$\times$ 10$^{-2}$ nm$^{-1}$ for the indicated temperatures.}
\label{fig:g2D2O} 
\end{figure}

Fig.\ref{fig:g2D2O} shows the typical behavior of the normalized intensity autocorrelation functions for an IPN sample at weight concentrations $C_w$=0.32 \% (Fig.\ref{fig:g2D2O}(a)) and $C_w$=0.10 \% (Fig.\ref{fig:g2D2O}(b)) and pH 7, collected at Q=1.8 $\times$ 10$^{-2}$ nm$^{-1}$ .

DLS measurements have been performed with a multiangle light
scattering setup. The monochromatic and polarized beam  emitted
from a solid state laser (100 mW at $\lambda$=642 nm) is focused on the sample placed in a cylindrical VAT for index matching and temperature control. The scattered
intensity is simultaneously collected at five different scattering
angles, namely $\theta$=30\textdegree, 50\textdegree, 70\textdegree,
90\textdegree, 110\textdegree, corresponding to five
scattering vectors Q (Q=0.67 $\times$ 10$^{-2}$ nm$^{-1}$, 1.1 $\times$ 10$^{-2}$  nm$^{-1}$, 1.5 $\times$ 10$^{-2}$ nm$^{-1}$, 1.8 $\times$ 10$^{-2}$ nm$^{-1}$, 2.1 $\times$ 10$^{-2}$ nm$^{-1}$), 
according to the relation Q=(4$\pi$n/$\lambda$) sin($\theta$/2). Single mode optical fibers
coupled to collimators collect the scattered light as a function
of time and scattering vector. In this way one can simultaneously measure 
the normalized intensity autocorrelation function
$g_2(Q,t)=<I(Q,t)I(Q,0)>/<I(Q,0)>^{2}$ at five different Q values
with a high coherence factor close to the ideal unit value.
Measurements have been performed as a function of temperature
in the range T=(293$\div$313) K across the VPT at four different weight concentrations ($C_w$=0.10 \%, $C_w$=0.15 \%, $C_w$=0.20 \%, $C_w$=0.32 \%), at both acidic and
neutral pH. Reproducibility has been tested by repeating
measurements several times. 

As commonly known, the intensity correlation function of most
colloidal systems is well described by the Kohlrausch-Williams-Watts
expression~\cite{KohlrauschAnnPhys1854, WilliamsFaradayTrans1970}:

\begin{equation}
g_2(Q,t)=1+b[(e^{-t/\tau})^{\beta}]^{2} \label{Eqfit}
\end{equation}

where $b$ is the coherence factor, $\tau$ is an "effective"
relaxation time and $\beta$ describes the deviation from the simple exponential decay ($\beta$
= 1) usually found in monodisperse systems and gives a measure of the distribution of relaxation times. Many glassy materials show a stretching of the correlation functions (here referred to as "stretched behavior") characterized by an exponent $\beta$
< 1.

\section{Results}
\label{Results}

\subsection{Isotopic effect on the dynamics}

\begin{figure*}[!t]
\centering
\includegraphics[height=8.5cm]{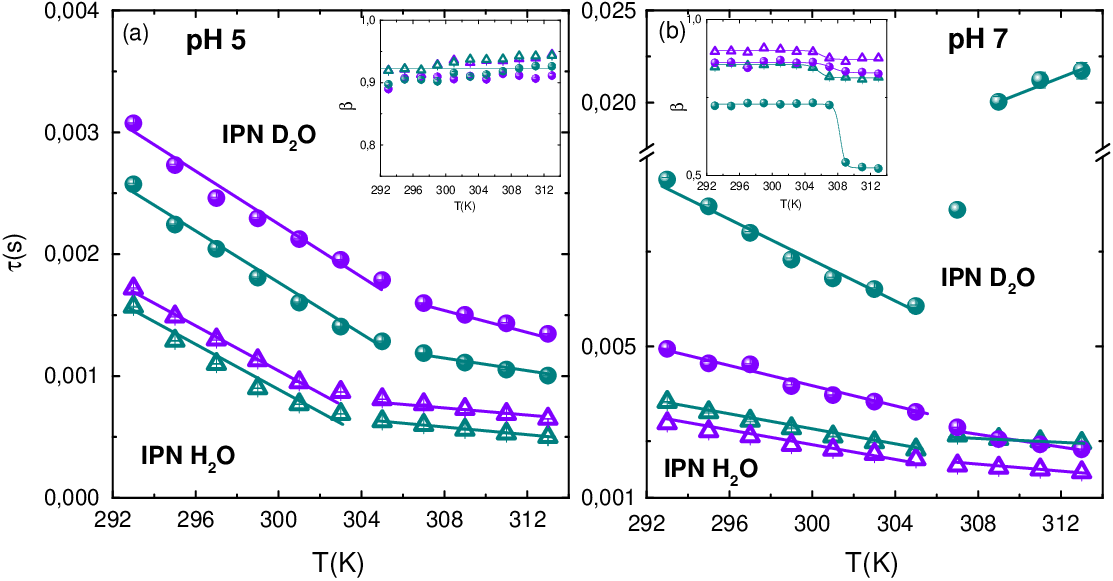}
\caption{Relaxation time and stretching parameter (inset) as a function of temperature for D$_2$O and H$_2$O suspensions of IPN microgels at $C_w$=0.15 \% (violet) and $C_w$=0.32 \% (cyan) at (a) pH 5 and (b) pH 7 for Q=1.8$\times$ 10$^{-2}$ nm$^{-1}$. Full lines are guides for eyes.}
\label{fig:tau2Cw}
\end{figure*}

In Fig.\ref{fig:tau2Cw} the comparison between the temperature dependence of the relaxation time and the $\beta$ parameter, as obtained through a fit with Eq.(\ref{Eqfit}), for D$_2$O and H$_2$O suspensions, at (a) pH 5 and (b) pH 7, is reported at two fixed concentrations as an example.
In D$_2$O a dynamical transition associated to the VPT, from a swollen to a shrunken state, is evidenced as for H$_2$O suspensions \cite{NigroJNCS2015}. At acidic pH (Fig.~\ref{fig:tau2Cw}(a)) the relaxation time slightly decreases as temperature increases, until the transition, evidenced by a change of the slope, is approached around T=305 K for H$_2$O and T=307 K for D$_2$O. Above this temperature the relaxation time decreases to its lowest value, corresponding to the shrunken state.
Moreover the stretching parameter $\beta$ (inset of Fig.~\ref{fig:tau2Cw}(a)) is almost constant with temperature and concentration and indicates slightly stretched correlation functions ($\beta \approx 0.9$).

This behavior is strongly affected by the pH of the solution, as shown in Fig.\ref{fig:tau2Cw}(b): at pH 7 the relaxation 
time slowly decreases as temperature increases up to the VPTT as for pH 5; above this temperature the relaxation time of low concentrated samples  ($C_w$=0.15 \%) decreases, whilst that of samples at high concentration ($C_w$=0.32 \%) increases. This effect is well enhanced in D$_2$O. 
At the same time the stretching parameter $\beta$, at variance with the case of pH 5, decreases with concentrations showing a step across the VPTT. As for the relaxation time at the highest concentration in D$_2$O $\beta$ shows a significant jump to lower values (inset of Fig.\ref{fig:tau2Cw}(b)), corresponding to a clear change of the shape of the intensity autocorrelation functions, well evidenced in Fig.\ref{fig:g2D2O}(a). Interestingly this peculiar behavior both for $\tau$ and $\beta$ can be interpreted as the first evidence of an aggregation phenomenon and as a precursor of the transition from an ergodic to a non-ergodic state expected at even higher concentrations.

Although the main features of the typical swelling behavior in H$_2$O suspensions \cite{NigroJNCS2015} are preserved under isotopic substitution, interesting differences between D$_2$O and H$_2$O samples are observed. As evidenced in Fig.\ref{fig:tau2Cw}, the relaxation times are always higher in D$_2$O than in H$_2$O, at all pH and concentrations, suggesting a slowing down of the dynamics under isotopic substitution, probably due to the higher viscosity of D$_2$O than H$_2$O. Moreover at pH 5 a shift of the VPTT to higher temperature with respect to H$_2$O is observed, in agreement with what found for PNIPAM microgel suspensions \cite{ShirotaMacromolSymp2004}. On the contrary, at pH 7 the VPT occurs at the same temperature in both solvents, confirming that pH affects the role played by H-bondings. 
Finally, at neutral pH and at the highest concentration the jump of the relaxation time to higher values above the VPTT, is greatly amplified under H/D isotopic substitution on the solvent, as evidenced in Fig.\ref{fig:tau2Cw}(b). The $\beta$ parameter is significantly affected by isotopic substitution only for the highest concentration at pH 7, as evident from its lower values and the jump across the VPTT.  
This may suggest that aggregation is promoted in D$_2$O solvent at neutral pH, since the slowing down of the dynamics and the reduction of the mutual interference between PNIPAM and PAAc networks drive the system to arrest.

\begin{figure}[t]
\centering
\includegraphics[height=7cm]{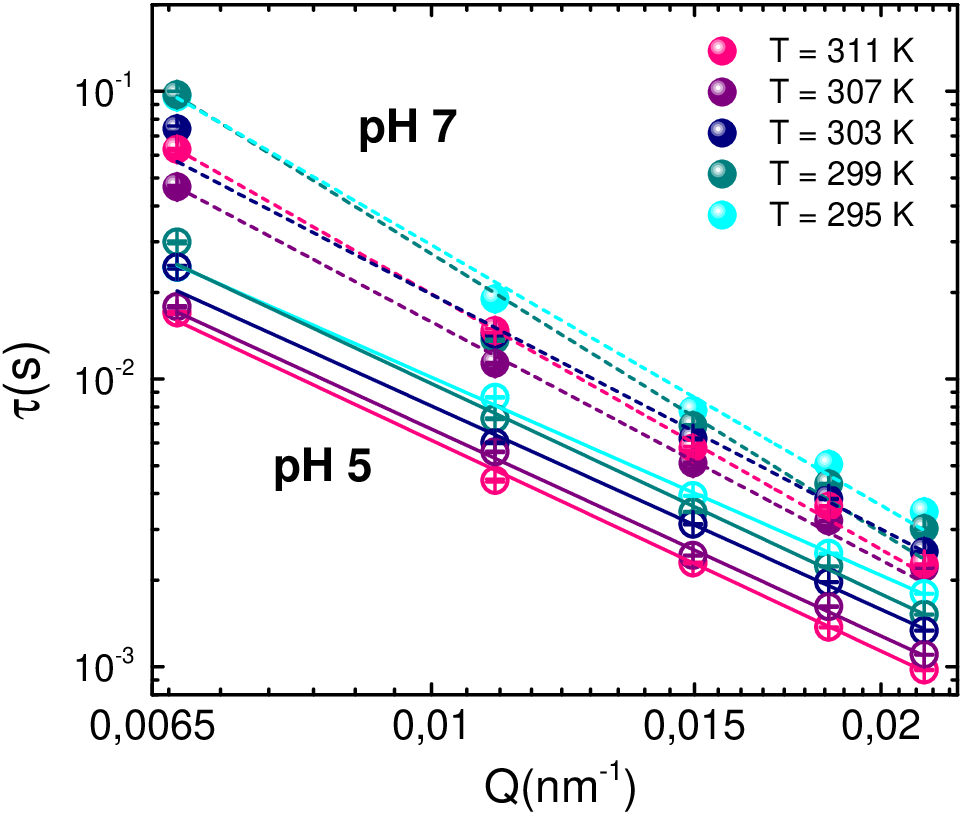}
\caption{Relaxation time as a function of
scattering vector Q at $C_w$=0.20 \% at pH 5 and pH 7 for the indicated temperatures. Full (pH 5) and dashed (pH 7) lines are fits through Eq.(\ref{EqPowerLaw}) with $\bar{n}_{pH5}$=2.3$\pm$0.1 and $\bar{n}_{pH7}$=2.9$\pm$0.1.}
\label{fig:tauQ D2O} 
\end{figure}

In order to obtain additional information on the dynamical behavior of D$_2$O IPN microgel  suspensions, the relaxation time and the stretching parameter have been investigated at different length scales, as a function of the scattering vector Q well below the peak position of the static structure factor at Q $\sim$ 0.08 nm$^{-1}$, as obtained from SAXS measurements in the Q-range $0.015 \div 0.5$ nm$^{-1}$ at T=(293$\div$313) K and with an incident X-ray energy fixed at 12.4 keV. In Fig.\ref{fig:tauQ D2O}, the Q-dependence of the relaxation time at $C_w$=0.20 \% at pH 5 and pH 7, are reported as an example. It can be described by a typical power law decay:

\begin{equation}
\tau=AQ^{-n} \label{EqPowerLaw}
\end{equation}

where $A$ is a constant and the exponent $n$ defines the nature of the motion.
The fits according to Eq.(\ref{EqPowerLaw}) are superimposed to the data as full or dashed lines in Fig.\ref{fig:tauQ D2O}. The mean exponents $\bar{n}$ are $\bar{n}_{pH5}$=2.3$\pm$0.1 and $\bar{n}_{pH7}$=2.9$\pm$0.1 in agreement with measurements on IPN microgels suspensions in H$_2$O \cite{NigroJNCS2015}. This peculiar behavior may be due to the complex morphology of IPN microgel particles,  with a highly cross-linked core and a less crosslinked corona, which may affect the dynamics. Moreover concentration and temperature do not significantly affect the Q dependence of the relaxation time. Similar results have already been published for different polymers~\cite{ColmeneroPRB1991, ColmeneroPRL1992} and responsive microgels \cite{ScherzingerPCCP2012}, although a theoretical explanation is not yet available. 
Meanwhile the stretching parameter $\beta$ does not show any dependence on the scattering vector Q as observed in H$_2$O suspensions \cite{NigroJNCS2015}.

\subsection{Swelling behavior and theoretical model}

\begin{figure*}[t]
\centering
\includegraphics[height=8.5cm]{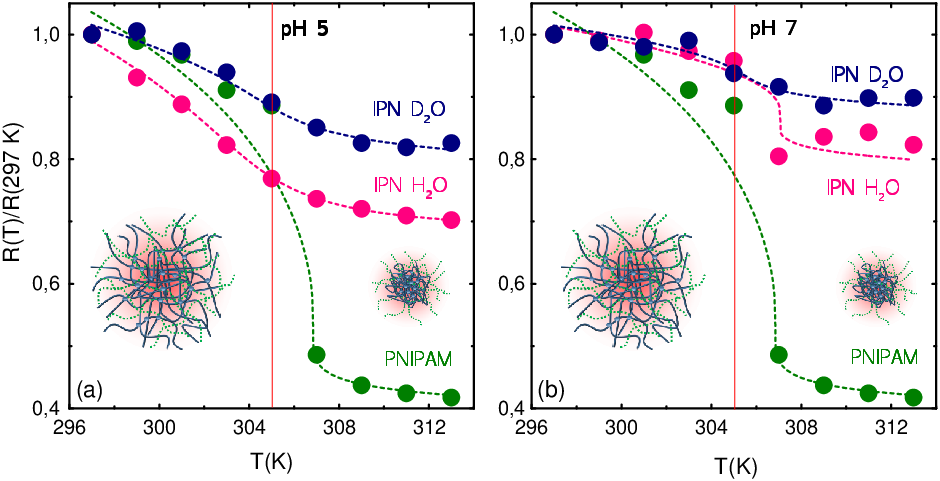}
\caption{Normalized radius as obtained from DLS
measurements for an IPN microgel in both H$_2$O and D$_2$O suspensions at $C_w=0.10$ \%,  Q=1.8$\times$ 10$^{-2}$ nm$^{-1}$ and (a) pH 5 and (b) pH 7, compared with the normalized radius obtained for PNIPAM microgels. The hydrodynamic radii have been normalized at the corresponding values at T=297 K: 
$R_{PNIPAM}=(55.1\pm 0.2)$ nm; 
$R_{IPN \; H_2O \; pH 5}=(97.7\pm 0.3)$ nm; $R_{IPN \; D_2O \; pH 5}=(95.7\pm 0.4)$ nm; $R_{IPN \; H_2O \; pH 7}=(135.7 \pm 0.5)$ nm and $R_{IPN \; D_2O \; pH 7}=(119.1\pm 0.3)$ nm. Dashed lines are the best fits through the modified Flory-Rehner theory as explained in the text.
}
\label{Rnorm}
\end{figure*}

The temperature dependence of the normalized hydrodynamic radii of IPN microgels in both H$_2$O and D$_2$O suspensions are shown in Fig.~\ref{Rnorm}(a) at acidic pH and in Fig.~\ref{Rnorm}(b) at neutral pH and compared with PNIPAM microgels at the same weight concentration ($C_w=0.10$ \%). They have been calculated, according to the Stokes-Einsteins relation for Brownian particles in the high dilution limit, as R=K$_B$ T /6 $\pi$ $\eta$ D, where D is the translational diffusion coefficient calculated from $\tau$ through the relation $\tau$=1/Q$^2$D. The sample viscosity $\eta$ has been approximated with the solvent one and the radii have been normalized with respect to their values at $T$=297 K reported in the caption of Fig.~\ref{Rnorm}. Fig.~\ref{Rnorm} evidences a clear reduction of the normalized radii of IPN microgels in both solvents compared to PNIPAM microgel, thus confirming that the presence of poly(acrylic acid) reduces the swelling capability of the microgel particles also in deuterated suspensions~\cite{HuAdvMater2004, XiaLangmuir2004, JonesMacromol2000}. Nevertheless different behaviors depending on solvent and pH are observed. For D$_2$O suspensions at acidic pH the range of variability of $R$ is reduced with respect to H$_2$O, while at neutral pH the transition is smoother than in H$_2$O. Moreover while in H$_2$O a clear difference is observed between the temperature behavior of the normalized radii at acidic and neutral pH, no significant changes are evident in D$_2$O. This suggests that the balance between polymer/polymer and polymer/solvent interactions strictly depends on the solvent and therefore on the H-bondings.

The swelling behavior of microgels has been widely described via the Flory-Rehner theory \cite{Flory1953} which gives the equation of state at the equilibrium condition

\begin{equation}
ln(1-\phi)+\phi+\chi \phi^2+\frac{\phi_0}{N}[(\frac{\phi}{\phi_0})^{1/3}-\frac{1}{2}\frac{\phi}{\phi_0}]=0
\label{EqState}
\end{equation}

\noindent $\phi$ is the polymer volume fraction within the particle and $\phi_0$ is the polymer volume fraction in the reference state, typically taken as the shrunken one. 
For isotropic swelling, $\phi$ can be related to the particle size $R$ as

\begin{equation}
\frac{\phi}{\phi_0}=(\frac{R_0}{R})^{3}
\label{phi}
\end{equation}

\noindent where $R_0$ is the particle diameter at the reference state and $R$ is the particle diameter at a given state. $N$ is the average number of segments between two neighboring cross-linking points in the gel network and $\chi$ is the Flory polymer-solvent energy parameter that can be written as a power series expansion

\begin{equation}
\chi=\chi_1 (T)+\chi_2 \phi+\chi_3 \phi^2+\chi_4 \phi^3+\cdots
\label{Chi3}
\end{equation}

where $\chi_1$ is the Flory parameter and $\chi_2,\chi_3,\chi_4,\ldots$ are temperature independent coefficients that introduce additional terms in the equation of state (Eq.\ref{EqState}). This model well describes the VPT in the case of pure PNIPAM microgels as shown in Fig.~\ref{Rnorm}(a)
where a second-order approximation of Eq.\ref{Chi3} is considered. However it does not reproduce the discontinuous transition observed in other soft colloidal microgels where exchange interactions must be taken into account considering higher order interactions than contacts between molecules~\cite{ErmanMacromol1986, LopezLeonPRE2007, NigroCSA2017}. Therefore for IPN microgels a careful choice of the order approximation of $\chi(\phi)$ is required. 
Indeed for IPN in H$_2$O  at acidic pH the best fit of our experimental data is obtained via a second-order approximation of Eq.\ref{Chi3}, such as for PNIPAM microgels. On the contrary, for H$_2$O samples at neutral pH and for D$_2$O samples at both acidic and neutral pH, a third-order approximation is needed for a good agreement between theory and experiments. This comparison has recently provided details about the delicate balance between energetic and entropic contribution assuming that $\chi$ is an effective mean parameter accounting for polymer/polymer interactions within each network, polymer/polymer interactions between different networks and polymer/solvent interactions.
We have indeed observed that the swelling process in H$_2$O is more favored at acidic than neutral pH, being the system more hydrophobic. Moreover it hugely depends on solvent: in D$_2$O the pH influence is less significant than in H$_2$O, confirming that the balance between polymer/polymer and polymer/solvent interactions strictly depends on the solvent and therefore on the H-bondings \cite{NigroCSA2017}. 

\subsection{Fragility in H$_2$O and D$_2$O suspensions of IPN microgels}

It is well known that in supercooled molecular liquids the structural relaxation time and the viscosity grow many order of magnitude with decreasing temperature when the glassy state is approached \cite{AngellPNAS1995}. The rapidity with which these quantities increase can be quantified by the fragility index $m$ defined as

\begin{equation} 
m= \left[\frac{\partial log \tau}{\partial (T_g/T)}\right]_{T=T_g} \label{fragilityT}
\end{equation}

where $\tau$ is the relaxation time and $T_g$ is the glass transition temperature.
In "fragile" liquids the relaxation time and viscosity are highly sensitive to changes in T with a super-Arrhenius (Vogel Fulcher Tammann) behavior, while in "strong" liquids they have a much lower T sensitivity with an Arrhenius dependence.
This unifying concept describes and classifies the behavior of glass forming systems \cite{AngellPNAS1995, AngellJAP2000, BohmerJCP1993}. The same concept has been extended to colloids where the glass transition occurs by increasing the volume fraction \cite{MattssonNature2009, CasaliniJCP2012}.
A recent work on IPN microgel suspensions in water at different concentrations and fixed temperature has related the fragility of soft colloidal particles to their elastic properties\cite{MattssonNature2009}.  

\begin{figure}[t]
\centering
\includegraphics[height=7cm]{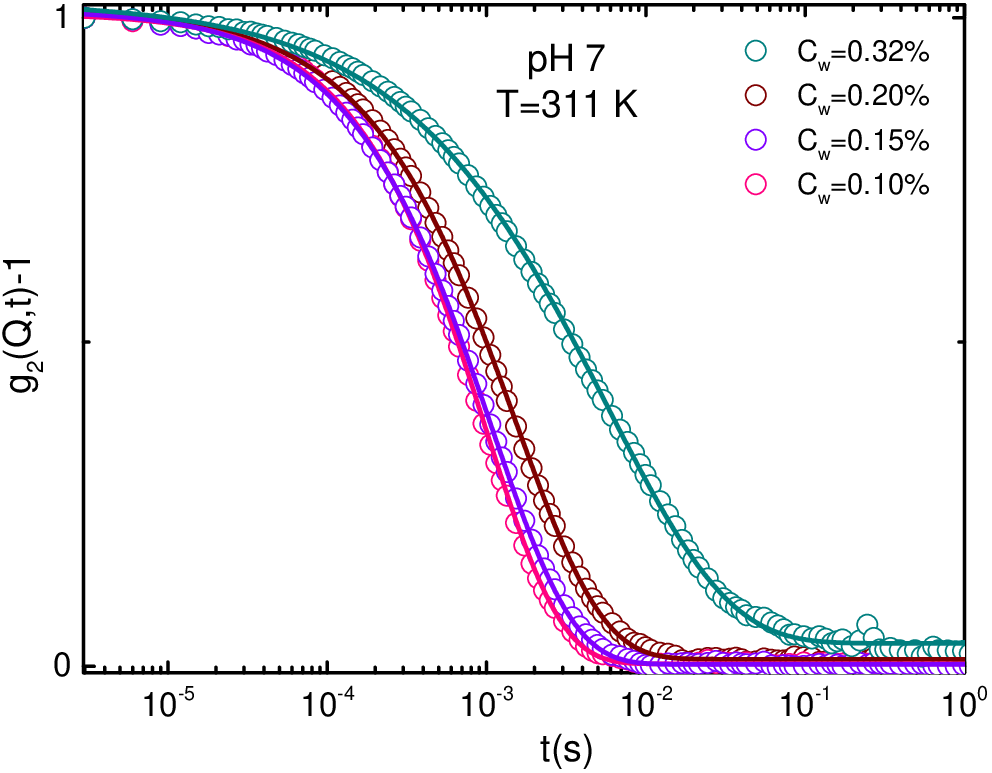}
\caption{Normalized intensity autocorrelation functions of a D$_2$O suspension of IPN microgels at fixed temperature (T=311 K), pH 7 and Q=1.8$\times$ 10$^{-2}$ nm$^{-1}$ for the indicated concentrations. Lines superimposed to data are fits according to Eq.(\ref{Eqfit}).}
\label{fig:g2Cw} 
\end{figure}

\begin{figure}[t]
\centering
\includegraphics[height=7cm]{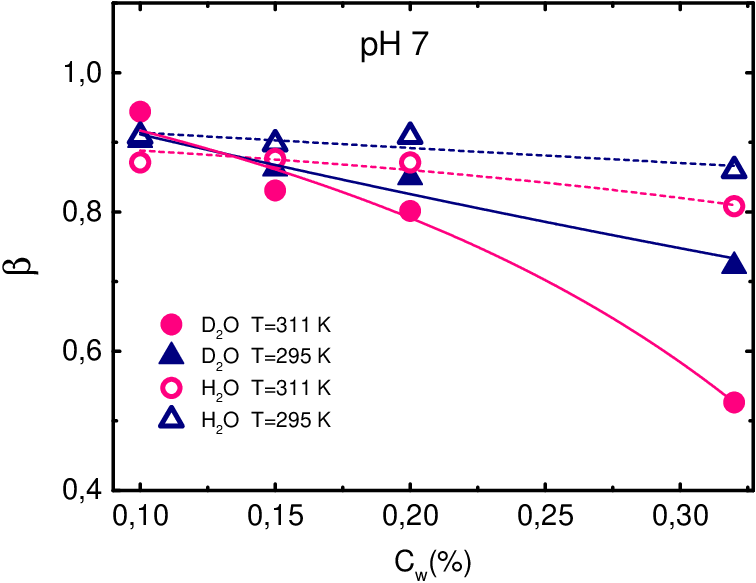}
\caption{Stretching coefficient as obtained from normalized intensity autocorrelation functions for D$_2$O and H$_2$O suspensions of IPN microgels at fixed temperature (T=311 K), pH 7 and Q=1.8$\times$ 10$^{-2}$ nm$^{-1}$ for the indicated concentrations. Full (D$_2$O) and dashed (H$_2$O) lines are Arrhenius fits for T<VPTT and Vogel-Fulcher-Tamman fits for T>VPTT.}
\label{fig:betaCw}
\end{figure}

In this work we perform an extensive investigation at different concentrations, temperatures and solvents.
The concentration dependence of the dynamics of IPN microgel suspensions is highlighted in Fig.\ref{fig:g2Cw}, where the normalized intensity autocorrelation functions at fixed temperature (T=311 K) and pH 7 are reported.
The observed behavior indicates that, as concentration increases, the decay becomes more and more stretched, 
as well evidenced in Fig.\ref{fig:betaCw}, where the concentration behavior of the stretching parameter $\beta$ at fixed temperatures (T=295 K and T=311 K) and pH 7, is reported for both D$_2$O and H$_2$O suspensions. 
We find that $\beta$ decreases with concentration showing an enhancement in the case of D$_2$O above the VPTT (T=311 K).

\begin{figure*}[t]
\centering
\includegraphics[height=8cm]{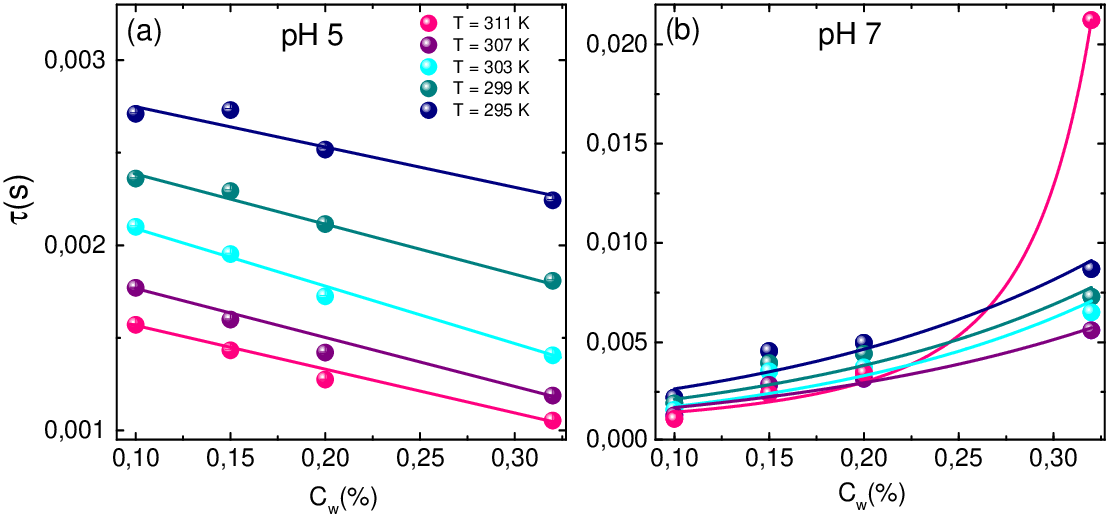}
\caption{Relaxation times as a function of weight concentration at fixed temperatures at (a) pH 5 and (b) pH 7 for D$_2$O suspensions at Q=1.8$\times$ 10$^{-2}$ nm$^{-1}$. Full lines in (a) are linear fits and full lines in (b) are fits according to Eq.\ref{Arrhenius} below the VPTT and Eq.\ref{VFT} above the VPTT.}
\label{fig:tauCwpH57}
\end{figure*}

The comparison between the concentration dependence of the relaxation time at acidic and neutral pH is reported in Fig.\ref{fig:tauCwpH57}(a) and (b). At pH 5 $\tau$ linearly decreases as concentration increases indicating a slight fastening of the dynamics while at pH 7 it exhibits an inverted trend with an exponential growth with different sensitivities to concentration below and above the VPT, suggesting the existence of two different routes to approach the glass transition. Moreover a faster rise at the highest temperature (T=311 K), indicates the appearance of an aggregation phenomenon. The different behavior observed in the investigated concentration range at pH 5 and pH 7 implies that the presence of PAAc, with respect to pure PNIPAM, hugely affects the dynamics of the system. 
The concentration decrease of the relaxation time at acidic pH is consistent with theoretical works that predict an increase with concentrations of the diffusion coefficient \cite{BanchioJCP2008, GapinskiJCP2009, GapinskiJCP2007} on charged colloidal particles. This indicates that the electrostatic effect plays an important role in the dynamics of the studied IPN microgels. New scenarios could appear at higher concentrations. The reversed  behavior observed at neutral pH can be related to intervening short-range attractive interactions due to hydrogen bonding between deprotonated carboxylic groups of PAAc  chains belonging to distinct particles.  Indeed, the ionization of PAAc at increasing pH corresponds to a transition from an insoluble to a soluble state of the polymeric chains and even if the interconnection of PAAc in the PNIPAM network avoid its diffusion out of particles, it can not avoid the interaction between the freely moving chain-end segments at the particles surface. A deeper investigation on the charge of the system is in progress, to shed light on the relevance of the electrostatic effect on the dynamics.

\begin{figure}[!t]
\centering
\includegraphics[height=8cm]{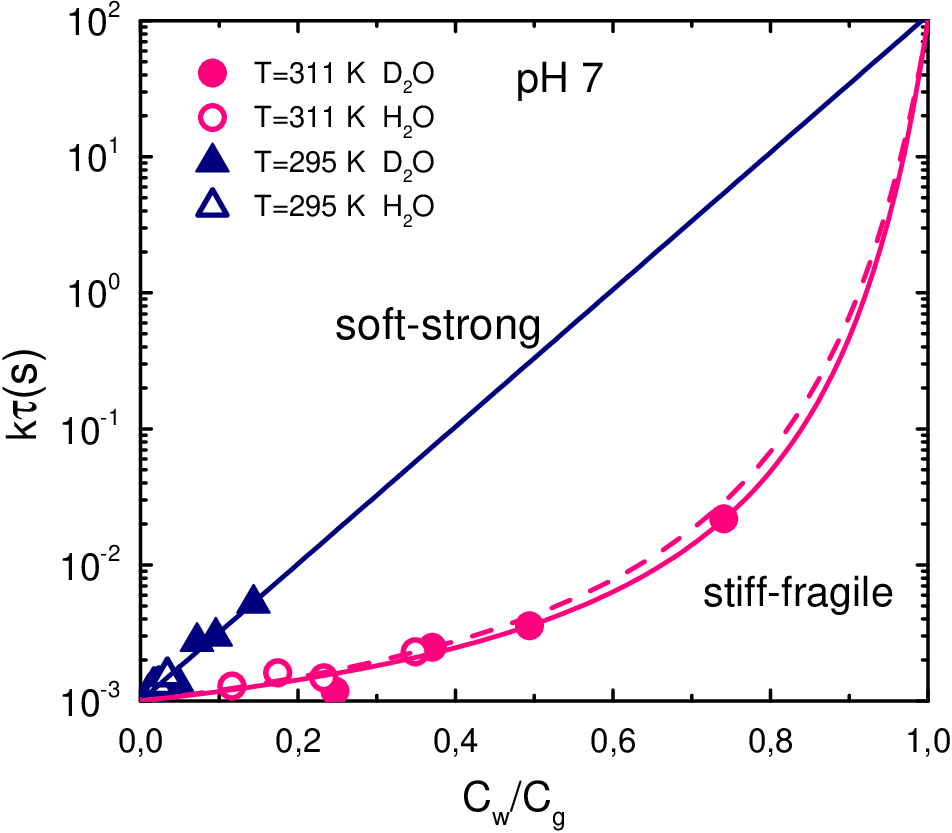}
\caption{Arrhenius plot for the scaled relaxation time $k\tau$ with $k=10^{-3}/\tau_0$ versus concentration $C_w$ normalized by $C_g=C_w(\tau$(100 s)) at T=295 K (triangles) and T=311 K (circles) for D$_2$O (solid symbols) and H$_2$O (open symbols). Lines superimposed to data are fits according to the Arrhenius (black line) and the VFT (magenta line) equations, for soft and stiff particles, respectively.}
\label{fig:APlot}
\end{figure}

Indeed we found that below the VPT the data are well fitted with an Arrhenius behavior, typically observed in strong molecular glass-formers, where $C_w$ plays the role of $1/T$. Therefore at these temperatures data are well described by the relation

\begin{equation} 
\tau=\tau_0\:e^{\;A C_w} \label{Arrhenius}
\end{equation}

where $A$ is a constant and $\tau_0$ is the characteristic relaxation time for low values of $C_w$.
On the other hand above the VPT, data are strongly dependent on concentration, cannot be described through a simple Arrhenius model and a super-Arrhenius behavior as the Vogel-Fulcher-Tammann (VFT) has to be considered:

\begin{equation} 
\tau=\tau_0 \: e^{\; \frac{\bar{A}C_w}{C_0-C_w}} \label{VFT}
\end{equation}

where $C_0$ sets the apparent divergence, $\bar{A}$ controls the growth of the relaxation time on
approaching $C_0$ and $\tau_0$ is the characteristic relaxation time at low concentrations. These functions provide a good description of the concentration dependence of $\tau$ and $\beta$ as shown in Fig.\ref{fig:betaCw} and Fig.\ref{fig:tauCwpH57}, thus confirming that the paradigm for supercooled
molecular liquids near their glass transition can be extended to suspensions of soft particles, where the concentration $C_w$ plays a role analogous to the inverse of temperature $1/T$ in molecular systems.

Generally the concept of fragility is summarized in a renormalized Arrhenius plot, where the temperature is rescaled by the glass transition temperature, $T_g$, and fragility is defined by the logarithmic slope at $T_g$ \cite{AngellPNAS1995}. 
This representation provides a unifying framework to describe the variation from strong to fragile behavior
of molecular liquids.
According to Mattsson et al. \cite{MattssonNature2009}, we explore the analogy between soft colloidal suspensions and molecular glass-formers by rescaling the Arrhenius plot in a fashion similar to that used for molecular glasses.
Indeed the corresponding plot for colloids can be obtained  by scaling $C_w$ by the "glass concentration" $C_g$, defined as the concentration $C_w(\tau$=100 s), where the structural relaxation time is no longer experimentally accessible \cite{AngellJAP2000}. In this way we obtain a renormalized Arrhenius plot for D$_2$O and H$_2$O suspensions, as shown in Fig.\ref{fig:APlot} at pH 7, where only two temperatures are reported since all data below the VPTT collapse on T=295 K and all data above collapse on T=311 K. The slope of the data at $C_w=C_g$ defines the fragility:

\begin{equation} 
m= \left[\frac{\partial log \tau}{\partial (C_w/C_g)}\right]_{C_w=C_g} \label{fragility}
\end{equation}
\\
As discussed before the most fragile materials are those that show the largest deviations from the Arrhenius law. In our case we observe that both in D$_2$O and H$_2$O for all temperatures below the VPT, in the swollen state, soft IPN microgels behave as strong materials. On the contrary above the VPT, in the shrunken state, they become stiff and behave as fragile systems. Moreover in this case the fragility of the system can be tuned by changing the solvent. In particular for all T<VPTT we find $m\approx 5$ in both H$_2$O and D$_2$O suspensions and for all T>VPTT, we find $m\approx30$ in H$_2$O and $m\approx 40$ in D$_2$O, where higher values of $m$ correspond to stiffer particles. This means that across the VPT the system encompasses a transition from a strong to a fragile behavior in both solvents. Nevertheless we find that above the VPT, microgel particles in D$_2$O suspensions are stiffer than in H$_2$O, thus leading to a more fragile behavior. Interestingly, this observation is reminiscent of what seen for the concentration dependence of $\beta$, where lower values correspond to stiffer particles and therefore to more fragile behaviors. At variance with previous studies, where the fragility of the particles was varied by changing the polymeric networks, here we are able to pass from a strong (soft) to fragile (stiff) behavior on a single system by changing temperature below and above the VPT.

These findings confirm that the fragility of colloidal particles increases as softness decreases, according to the experimental results by Mattsson et al. \cite{MattssonNature2009}. Moreover our study is extended to different temperatures and solvents providing the possibility to modulate the inherent softness of microgel particles and therefore their fragility. This point, still controversial, calls for further experimental,  numerical and theoretical investigations.

\section{Conclusions}
\label{Conclusions}

The swelling behavior of PNIPAM-PAAc IPN microgels has been investigated as a function of temperature, solvent, pH, concentration and scattering vector by comparing new experiments on deuterated suspensions with previous findings on IPN microgels in H$_2$O \cite{NigroJNCS2015}.

It has been shown that H/D isotopic substitution in the solvent plays an important role on the kinetics of the swelling although preserving the same physical properties observed in H$_2$O. This suggests that hydrogen-bonds play a crucial role in the polymer-solvent interactions and that the swelling kinetics can be slightly affected by isotopic substitution. In this case, indeed, stronger H-bondings occur between polymer and solvent with respect to the case of H$_2$O with consequent changes in the rate of the swelling/shrinking transition \cite{ShirotaMacromolSymp2004}.

Moreover the dynamics of IPN microgels is slowed down in D$_2$O solvent.
Importantly for the highest concentration at pH 7 a clear enhanced transition of the relaxation time above the VPTT compared to H$_2$O is found. This behavior can be interpreted as a precursor
of a non-ergodic transition  expected at even higher concentrations.

Experimental data have been compared with theoretical models from the Flory-Rehner theory: while  PNIPAM microgels data are well described through the second order approximation of the Flory parameter $\chi (\phi)$, for IPN microgels 
at neutral conditions and in D$_2$O solvents we extended the model by 
considering a third-order contribution. 

Finally, applying to colloidal systems the universal framework of fragility \cite{MattssonNature2009}, we find that the elastic response of IPN microgel suspensions can be modulated by changing temperature: below the VPT particles are soft and deformable, whilst above the VPT they become stiff and undeformable. This is associated to increasing values of fragility, which suggests that the swelling/shriking behavior drives the system through a transition from strong to fragile.  Therefore fragility has to be strictly related to the VPT: soft particles will lead to strong behavior and stiff particles to fragile behavior.

\section*{Acknowledgments}

The authors acknowledge support from the European Research Council (ERC Consolidator Grant
681597, MIMIC) and from MIUR-PRIN (2012J8X57P).

%%%REFERENCES%%%

%\bibliography{bib_16-05-17}

\begin{thebibliography}{10}
\bibliographystyle{unsrt}\biboptions{sort&compress}

\bibitem{SciAdvPhys2005}
F.~{Sciortino} and P.~{Tartaglia}.
\newblock {Glassy Colloidal Systems}.
\newblock {\em Adv. Phys.}, 54:471--524, 2005.

\bibitem{TrappeCOCIS2004}
V.~Trappe and P.~Sandk{\"u}hler.
\newblock {Colloidal gels - low-density disordered solid-like states}.
\newblock {\em Curr. Opin. Colloid Interface Sci.}, 8:494--500, 2004.

\bibitem{PoonCOCIS1998}
W.~C.~K. Poon.
\newblock {Phase separation, aggregation and gelation in colloid polymer
  mixtures and related systems}.
\newblock {\em Curr. Opin. Colloid Interface Sci.}, 3:593--599, 1998.

\bibitem{ZacJPCM2008}
E.~Zaccarelli.
\newblock Gelation as arrested phase separation in short-ranged attractive
  colloid-polymer mixture.
\newblock {\em J. Phys.: Condens. Matter}, 20:494242, 2008.

\bibitem{LuNat2008}
P.~J. Lu, E.~Zaccarelli, F.~Ciulla, A.~B. Schofield, F.~Sciortino, and D.~A.
  Weitz.
\newblock Gelation of particle with short range attraction.
\newblock {\em Nature}, 453:499--503, 2008.

\bibitem{RoyallNatMat2008}
C.~P. Royall, S.~R. Williams, T.~Ohtsuka, and H.~Tanaka.
\newblock {Direct observation of a local structural mechanism for dynamical
  arrest}.
\newblock {\em Nat. Mater.}, 7:556--561, 2008.

\bibitem{RuzickaNatMat2011}
B.~Ruzicka, E.~Zaccarelli, L.~Zulian, R.~Angelini, M.~Sztucki, A.~Moussa{\"i}d,
  T.~Narayanan, and F.~Sciortino.
\newblock {Observation of empty liquids and equilibrium gels in a colloidal
  clay}.
\newblock {\em Nat. Mater.}, 10:56--60, 2011.

\bibitem{PuseyNat1986}
P.~N. Pusey and W.~van Megen.
\newblock Phase behaviour of concentrated suspensions of nearly hard colloidal
  spheres.
\newblock {\em Nature}, 320:340--342, 1986.

\bibitem{ImhofPRL1995}
A.~Imhof and J.~K.~G. Dhont.
\newblock {Experimental Phase Diagram of a Binary Colloidal Hard-Sphere Mixture
  with a Large Size Ratio}.
\newblock {\em Phys. Rev. Lett.}, 75:1662--1665, 1995.

\bibitem{PhamScience2002}
K.~N. Pham, A.~M. Puertas, J.~Bergenholtz, S.~U. Egelhaaf, A.~Moussa{\"i}d,
  P.~N. Pusey, A.~B. Schofield, M.~E. Cates, M.~Fuchs, and W.~C.~K. Poon.
\newblock {Multiple Glassy States in a Simple Model System}.
\newblock {\em Science}, 296:104--106, 2002.

\bibitem{EckertPRL2002}
T.~Eckert and E.~Bartsch.
\newblock {Re-entrant glass transition in a colloid-polymer mixture with
  depletion attractions}.
\newblock {\em Phys. Rev. Lett.}, 89:125701--4, 2002.

\bibitem{AngeliniNC2014}
R.~Angelini, E.~Zaccarelli, F.~A. de~Melo~Marques, M.~Sztucki, A.~Fluerasu,
  G.~Ruocco, and B.~Ruzicka.
\newblock {Glass-glass transition during aging of a colloidal clay}.
\newblock {\em Nat. Commun.}, 5:4049, 2014.

\bibitem{LikosJPCM2002}
C.~N. Likos, N.~Hoffmann, H.~L{\"{o}}wen, and A.~A. Louis.
\newblock {Exotic fluids and crystals of soft polymeric colloids}.
\newblock {\em J. Phys. Cond. Matter}, 14:7681--7698, 2002.

\bibitem{RamirezJPCM2009}
P.~E. Ram{\'i}rez-Gonz{\'a}lez and M.~Medina-Noyola.
\newblock {Glass transition in soft-sphere dispersions}.
\newblock {\em J. Phys. Cond. Matter}, 21:075101--13, 2009.

\bibitem{MattssonNature2009}
J.~Mattsson, H.~M. Wyss, A.~Fernandez-Nieves, K.~Miyazaki, Z.~Hu, D.~Reichman,
  and D.~A. Weitz.
\newblock {Soft colloids make strong glasses}.
\newblock {\em Nature}, 462(5):83\textendash{}86, 2009.

\bibitem{AngellPNAS1995}
C.A. Angell.
\newblock {The old problems of glass and the glass transition, and the many new
  twists}.
\newblock {\em Proc. Natl. Acad. Sci. USA}, 92:6675--6682, 1995.

\bibitem{AngellJAP2000}
C.A. Angell, K.~L. Ngai, G.~B. McKenna, P.~F. McMillan, and S.~W. Martin.
\newblock {Relaxation in glass forming liquids and amorphous solids}.
\newblock {\em J. Appl. Phys.}, 88:3113--3157, 2000.

\bibitem{CasaliniJCP2012}
R.~Casalini.
\newblock {The fragility of liquids and colloids and its relation to the
  softness of the potential}.
\newblock {\em J. Chem. Phys.}, 137:204904--4, 2012.

\bibitem{McKennaPRE2016}
X.~Peng and G.~B. McKenna.
\newblock {Physical aging and structural recovery in a colloidal glass
  subjected to volume-fraction jump conditions}.
\newblock {\em Phys. Rev. E}, 93:042603--13, 2016.

\bibitem{SeekellSM2015}
R.~P. Seekell, P.~S. Sarangapani, Z.~Zhangb, and Y.~Zhu.
\newblock {Relationship between particle elasticity, glass fragility, and
  structural relaxation in dense microgel suspensions}.
\newblock {\em Soft Matter}, 11:5485--5491, 2015.

\bibitem{BordatPRL2004}
P.~Bordat, F.~Affouard, and M.~Descamps.
\newblock {Does the Interaction Potential Determine Both the Fragility of a
  Liquid and theVibrational Properties of Its Glassy State?}
\newblock {\em Phys. Rev. Lett.}, 93(10):105502--4, 2004.

\bibitem{SenguptaJCP2011}
S.~Sengupta, F.~Vasconcelos, F.~Affouard, and S.~Sastry.
\newblock {Dependence of the fragility of a glass former on the softness of
  interparticle interactions}.
\newblock {\em J. Chem. Phys.}, 135:194503--9, 2011.

\bibitem{ShiJCP2011}
Z.~Shi, P.~G. Debenedetti, F.~H. Stillinger, and P.~Ginart.
\newblock {Structure, dynamics, and thermodynamics of a family of potentials
  with tunable softness}.
\newblock {\em J. Chem. Phys.}, 135:084153--9, 2011.

\bibitem{McKennaJCP2014}
X.~Di, X.~Peng, and G.~B. McKenna.
\newblock {Dynamics of a thermo-responsive microgel colloid near to the glass
  transition}.
\newblock {\em J. Chem. Phys.}, 140:054903--9, 2014.

\bibitem{WangChemPhys2014}
H.~Wang, X.~Wu, Z.~Zhu, C.~S. Liu, and Z.~Zhang.
\newblock {Revisit to phase diagram of poly(N-isopropylacrylamide) microgel
  suspensions by mechanical spectroscopy}.
\newblock {\em J. Chem. Phys.}, 140:024908--6, 2014.

\bibitem{PritiJCP2014}
P.~S. Mohanty, D.~Paloli, J.~J. Crassous, E.~Zaccarelli, and P.~Schurtenberger.
\newblock {Effective interactions between soft-repulsive colloids: Experiments,
  theory and simulations}.
\newblock {\em J. Chem. Phys.}, 140:094901--9, 2014.

\bibitem{HellwegCPS2000}
T.~Hellweg, C.D. Dewhurst, E.~Br{\"u}ckner, K.Kratz, and W.Eimer.
\newblock {Colloidal crystals made of poly(N-isopropylacrylamide) microgel
  particles}.
\newblock {\em Colloid. Polym. Sci.}, 278:972--978, 2000.

\bibitem{PaloliSM2012}
D.~Paloli, P.~S. Mohanty, J.~J. Crassous, E.~Zaccarelli, and P.~Schurtenberger.
\newblock {Fluid\textendash{}solid transitions in soft-repulsive colloids}.
\newblock {\em Soft Matter}, 9:3000--3004, 2013.

\bibitem{WuPRL2003}
J.~Wu, B.~Zhou, and Z.~Hu.
\newblock {Phase behavior of thermally responsive microgel colloids}.
\newblock {\em Phys. Rev. Lett.}, 90(4):048304--4, 2003.

\bibitem{PeltonColloids1986}
R.~H. Pelton and Chibante P.
\newblock {Preparation of aqueous lattices with N-isopropylacrylamide}.
\newblock {\em Colloids Surf.}, 20:247\textendash{}256, 1986.

\bibitem{SaundersACIS1999}
B.~R. Saunders and B.~Vincent.
\newblock {Microgels particles as model colloids: theory, properties and
  applications}.
\newblock {\em Adv. Colloid Interface Sci.}, 80:1\textendash{}25, 1999.

\bibitem{PeltonAdvColloid2000}
R.~H. Pelton.
\newblock {Temperature-sensitive aqueous microgels}.
\newblock {\em Adv. Colloid Interface Sci.}, 85:1\textendash{}33, 2000.

\bibitem{DasAnnRevMR2006}
M.~Das, H.~Zhang, and E.~Kumacheva.
\newblock {MICROGELS: Old Materials with New Applications}.
\newblock {\em Annu. Rev. Mater. Res.}, 36:117\textendash{}142, 2006.

\bibitem{KargCOCIS2009}
M.~Karg and T.~Hellweg.
\newblock {New ``smart'' poly(NIPAM) microgels and nanoparticle microgel
  hybrids: Properties and advances in characterisation}.
\newblock {\em Curr. Opin. Colloid Interface Sci.}, 14:438--450, 2009.

\bibitem{LuProgPolSci2011}
Y.~Lu and M.~Ballauff.
\newblock {Thermosensitive core-shell microgels: From colloidal model systems
  to nanoreactors}.
\newblock {\em Prog. Polym. Sci.}, 36:767\textendash{}792, 2011.

\bibitem{WuMacromol2003}
J.~Wu, G.~Huang, and Z.~Hu.
\newblock {Interparticle Potential and the Phase Behavior of
  Temperature-Sensitive Microgel Dispersions}.
\newblock {\em Macromolecules}, 36:440\textendash{}448, 2003.

\bibitem{LyonRevPC2012}
L.~A. Lyon and A.~Fernandez-Nieves.
\newblock {The Polymer/Colloid Duality of Microgel Suspensions}.
\newblock {\em Annu. Rev. Phys. Chem.}, 63:25--43, 2012.

\bibitem{TanPolymers2010}
B.~H. Tan, R.~H. Pelton, and K.~C. Tam.
\newblock {Microstructure and rheological properties of thermo-responsive
  poly(N-isopropylacrilamide) microgels}.
\newblock {\em Polymers}, 51:3238\textendash{}3243, 2010.

\bibitem{ZhuMacroChemPhys1999}
P.~W. Zhu and D.~H. Napper.
\newblock {Light scattering studies of poly(N-isopropylacrylamide) microgel
  particle in mixed water-acetic acid solvents}.
\newblock {\em Macromol. Chem. Phys.}, 200:1950\textendash{}1955, 1999.

\bibitem{KratzBerBunsenges11998}
K.~Kratz and W.~Eimer.
\newblock {Swelling Properties of Colloidal Poly(N-Isopropylacrylamide)
  Microgels in Solution}.
\newblock {\em Ber. Bunsenges. Phys. Chem.}, 102:848\textendash{}854, 1998.

\bibitem{KratzPolymer2001}
K.~Kratz, T.~Hellweg, and W.~Eimer.
\newblock {Structural changes in PNIPAM microgel particles as seen by SANS, DLS
  and EM techniques}.
\newblock {\em Polymer}, 42:6631\textendash{}6639, 2001.

\bibitem{BaoMacromol2006}
L.~Bao and L.~Zhaj.
\newblock {Preparation of Poly(N-isopropylacrylamide) Microgels using Different
  Initiators Under Various pH Values}.
\newblock {\em Macromol. Sci.}, 43:1765\textendash{}1771, 2006.

\bibitem{HellwegLangmuir2004}
T.~Hellweg, C.~D. Dewhurst, W.~Eimer, and K.~Kratz.
\newblock {PNIPAM-co-polystyrene core-shell microgels: structure, swelling
  behavior, and crystallization.}
\newblock {\em Langmuir}, 20:4333\textendash{}4335, 2004.

\bibitem{HuAdvMater2004}
Z.~Hu and X.~Xia.
\newblock {Hydrogel nanoparticle dispersions with inverse thermoreversible
  gelation}.
\newblock {\em Adv. Mater.}, 16(4):305\textendash{}309, 2004.

\bibitem{MaColloidInt2010}
J.~Ma, B.~Fan, B.~Liang, and J.~Xu.
\newblock {Synthesis and characterization of
  Poly(N-isopropylacrylamide)/Poly(acrylic acid) semi-IPN nanocomposite
  microgels}.
\newblock {\em J. Colloid Interface Sci.}, 341:88\textendash{}93, 2010.

\bibitem{KratzColloids2000}
K.~Kratz, T.~Hellweg, and W.~Eimer.
\newblock {Influence of charge density on the swelling of colloidal
  poly(N-isopropylacrylamide-co-acrylic acid) microgels}.
\newblock {\em Colloids Surf. A}, 170:137\textendash{}149, 2000.

\bibitem{KratzBerBunsenges21998}
K.~Kratz, T.~Hellweg, and W.~Eimer.
\newblock {Effect of connectivity and charge density on the swelling and local
  structure and dynamic properties of colloidal PNIPAM microgels}.
\newblock {\em Ber. Bunsenges. Phys. Chem.}, 102(11):1603\textendash{}1608,
  1998.

\bibitem{XiaLangmuir2004}
X.~Xia and Z.~Hu.
\newblock {Synthesis and Light Scattering Study of Microgels with
  Interpenetrating Polymer Networks}.
\newblock {\em Langmuir}, 20:2094\textendash{}2098, 2004.

\bibitem{JonesMacromol2000}
C.~D. Jones and L.~A. Lyon.
\newblock {Synthesis and Characterization of Multiresponsive Core-Shell
  Microgels}.
\newblock {\em Macromolecules}, 33:8301\textendash{}8303, 2000.

\bibitem{NigroJNCS2015}
V.~Nigro, R.~Angelini, M.~Bertoldo, V.~Castelvetro, G.~Ruocco, and B.~Ruzicka.
\newblock {Dynamic light scattering study of temperature and pH sensitive
  colloidal microgels}.
\newblock {\em J. Non-Cryst. Solids}, 407:361 -- 366, 2015.

\bibitem{NigroJCP2015}
V.~Nigro, R.~Angelini, M.~Bertoldo, F.~Bruni, M.A. Ricci, and B.~Ruzicka.
\newblock Local structure of temperature and ph-sensitive colloidal microgels.
\newblock {\em J. Chem. Phys.}, 143:114904--8, 2015.

\bibitem{XiongColloidSurf2011}
W.~Xiong, X.~Gao, Y.~Zao, H.~Xu, and X.~Yang.
\newblock {The dual temperature/pH-sensitive multiphase behavior of
  poly(Nisopropylacrylamide-co-acrylic acid) microgels for potential
  application in \textit{in situ} gelling system}.
\newblock {\em Colloids Surf. B: Biointerfaces}, 84:103\textendash{}110, 2011.

\bibitem{MengPhysChem2007}
Z.~Meng, J.~K. Cho, S.~Debord, V.~Breedveld, and L.~A. Lyon.
\newblock {Crystallization Behavior of Soft, Attractive Microgels}.
\newblock {\em J. Phys. Chem. B}, 111:6992\textendash{}6997, 2007.

\bibitem{LyonJPCB2004}
L.~A. Lyon, J.~D. Debord, S.~B. Debord, C.~D. Jones, J.~G. McGrath, and M.~J.
  Serpe.
\newblock {Microgel Colloidal Crystals}.
\newblock {\em J. Phys. Chem. B}, 108:19099\textendash{}19108, 2004.

\bibitem{HolmqvistPRL2012}
P.~Holmqvist, P.~S. Mohanty, G.~N{\"a}gele, P.~Schurtenberger, and M.~Heinen.
\newblock {Structure and Dynamics of Loosely Cross-Linked Ionic Microgel
  Dispersions in the Fluid Regime}.
\newblock {\em Phys. Rev. Lett.}, 109:048302--5, 2012.

\bibitem{DebordJPCB2003}
S.~B. Debord and L.~A. Lyon.
\newblock {Influence of Particle Volume Fraction on Packing in Responsive
  Hydrogel Colloidal Crystals}.
\newblock {\em J. Phys. Chem. B}, 107:2927\textendash{}2932, 2003.

\bibitem{RomeoSM2013}
G.~Romeo and M.~Pica Ciamarra.
\newblock {Elasticity of compressed microgel suspensions}.
\newblock {\em Soft Matter}, 9:5401--5406, 2013.

\bibitem{RomeoJCP2012}
G.~Romeo, L.~Imperiali, J.W. Kim, A.~Fern\'{a}ndez-Nieves, and D.~A. Weitz.
\newblock {Origin of de-swelling and dynamics of dense ionic microgel
  suspensions}.
\newblock {\em J. Chem. Phys.}, 136:124905, 2012.

\bibitem{XiaJCRel2005}
X.~Xia, Z.~Hua, and M.~Marquez.
\newblock {Physically bonded nanoparticle networks: a novel drug delivery
  system}.
\newblock {\em J. Control. Release}, 103:21\textendash{}30, 2005.

\bibitem{ZhouBio2008}
J.~Zhou, G.~Wang, L.~Zou, L.~Tang, M.~Marquez, and Z.~Hu.
\newblock {Viscoelastic Behavior and In Vivo Release Study of Microgel
  Dispersions with Inverse Thermoreversible Gelation}.
\newblock {\em Biomacromolecules}, 9:142\textendash{}148, 2008.

\bibitem{XingCollPolym2010}
Z.~Xing, C.~Wang, J.~Yan, L.~Zhang, L.~Li, and L.~Zha.
\newblock {pH/temperature dual stimuli-responsive microcapsules with
  interpenetrating polymer network structure}.
\newblock {\em Colloid Polym. Sci.}, 288:1723\textendash{}1729, 2010.

\bibitem{LiuPolymers2012}
X.~Liu, H.~Guo, and L.~Zha.
\newblock {Study of pH/temperature dual stimuli-responsive nanogels with
  interpenetrating polymer network structure}.
\newblock {\em Polymers}, 61(7):1144\textendash{}1150, 2012.

\bibitem{SibandMacromol2011}
E.~Siband, Y.~Tran, and D.~Hourdet.
\newblock {Thermoresponsive Interpolyelectrolyte Complexation: Application to
  Macromolecular Assemblies}.
\newblock {\em Macromolecules}, 44:8185--8194, 2011.

\bibitem{NigroCSA2017}
V.~Nigro, R.~Angelini, M.~Bertoldo, and B.~Ruzicka.
\newblock Swelling behavior in multi-responsive microgels.
\newblock {\em Colloids Surf. A}, xxx:xxx--xxx
  https://doi.org/10.1016/j.colsurfa.2017.04.059, 2017.

\bibitem{ShirotaMacromolSymp2004}
H.~Shirota and K.~Horie.
\newblock {Deuterium Substitution and Fluorescence Studies on Polymer Hydrogels
  and Complexes}.
\newblock {\em Macromol. Symp.}, 207:79--93, 2004.

\bibitem{TudiscaRSC2012}
V.~Tudisca, M.A. Ricci, R.~Angelini, and B.~Ruzicka.
\newblock {Isotopic effect on the aging dynamics of a charged colloidal
  system}.
\newblock {\em RSC Advances}, 2:11111, 2012.

\bibitem{MarquesSM2015}
F.~A. de~Melo~Marques, R.~Angelini, E.~Zaccarelli, B.~Farago, B.~Ruta,
  G.~Ruocco, and B.~Ruzicka.
\newblock {Structural and microscopic relaxations in a colloidal glass}.
\newblock {\em Soft Matter}, 11:466--471, 2015.

\bibitem{MarquesJPC2017}
F.~A. de~Melo~Marques, R.~Angelini, G.~Ruocco, and B.~Ruzicka.
\newblock {Isotopic Effect on the Gel and Glass Formation of a Charged
  Colloidal Clay: Laponite}.
\newblock {\em J. Phys. Chem. B.}, 121:4576--4582, 2017.

\bibitem{KohlrauschAnnPhys1854}
R.~Kohlrausch.
\newblock {Thermoresponsive poly-(N-isopropylmethacrylamide) microgels:
  Tailoring particle size by interfacial tension control}.
\newblock {\em Pogg. Ann. Phys. Chem.}, 91:179\textendash{}214, 1854.

\bibitem{WilliamsFaradayTrans1970}
G.~Williams and D.~C. Watts.
\newblock {Non-Symmetrical Dielectric Relaxation Behavior Arising from a Simple
  Empirical Decay Function}.
\newblock {\em J. Chem. Soc. Faraday Trans.}, 66:80\textendash{}85, 1970.

\bibitem{ColmeneroPRB1991}
J.~Colmenero, A.~Alegr{\'{i}}a, J.~M. Alberdi, F.~Alvarez, and B.~Frick.
\newblock {Dynamics of the $\alpha$ relaxation of a glass-forming polymeric
  system: Dielectric, mechanical, nuclear-magnetic-resonance, and neutron
  scattering studies}.
\newblock {\em Phys. Rev. B}, 44:7321\textendash{}7329, 1991.

\bibitem{ColmeneroPRL1992}
J.~Colmenero, A.~Alegr{\'{i}}a, and A.~Arbe.
\newblock {Correlation between Non-Debye behavior and $Q$-behavior of the
  $\alpha$-relaxation in glass-forming polymeric systems}.
\newblock {\em Phys. Rev. Lett.}, 69:478\textendash{}481, 1992.

\bibitem{ScherzingerPCCP2012}
C.~Scherzinger, O.~Holderer, D.~Richter, and W.~Richtering.
\newblock {Polymer dynamics in responsive microgels: influence of cononsolvency
  and microgel architecture}.
\newblock {\em Phys. Chem. Chem. Phys.}, 14:2762\textendash{}2768, 2012.

\bibitem{Flory1953}
P.J. Flory.
\newblock {\em Principles of Polymer Chemistry}.
\newblock Cornell University, Ithaca, New York, 1953.

\bibitem{ErmanMacromol1986}
B.Erman and P.J.Flory.
\newblock Critical phenomena and transitions in swollen polymer networks and in
  linear macromolecules.
\newblock {\em Macromolecules}, 19:2342--2353, 1986.

\bibitem{LopezLeonPRE2007}
T.~L\`{o}pez-Le\`{o}n and A.~Fernandez-Nieves.
\newblock Macroscopically probing the entropic influence of ions: deswelling
  neutral microgels with salt.
\newblock {\em Phys. Rev. E}, 75:1--7, 2007.

\bibitem{BohmerJCP1993}
R.~B{\"{o}}hmer, K.~L. Ngai, C.A. Angell, , and D.J. Plazek.
\newblock {Nonexponential relaxations in strong and fragile glass formers}.
\newblock {\em J. Chem. Phys.}, 99(5):4201--9, 1993.

\bibitem{BanchioJCP2008}
{A. J. Banchio and G. Nag\"{a}gele}.
\newblock {Short-time transport properties in dense suspensions: From neutral
  to charge-stabilized colloidal spheres}.
\newblock {\em {J. Chem. Phys}}, {128}:{104903}, {2008}.

\bibitem{GapinskiJCP2009}
{J. Gapinski and A. Patkowski and A. J. Banchio and J. Buitenhuis and P.
  Holmqvist and M. P. Lettinga and G. Meier and G. Nag\"{a}gele}.
\newblock {Structure and short-time dynamics in suspensions of charged silica
  spheres in the entire fluid regime}.
\newblock {\em J. Chem. Phys}, 130:084503, 2009.

\bibitem{GapinskiJCP2007}
{J. Gapinski and A. Patkowski and A. J. Banchio and P. Holmqvist and G. Meier
  and M. P. Lettinga and G. Nag\"{a}gele}.
\newblock {Collective diffusion in charge-stabilized suspensions: Concentration
  and salt effects}.
\newblock {\em J. Chem. Phys}, 126:104905, 2007.

\end{thebibliography}
%\bibliographystyle{unsrt}\biboptions{sort&compress}

\end{document}